\documentclass[journal=jacsat,manuscript=article]{achemso}

\usepackage[version=3]{mhchem} 
\usepackage{xcolor}
\usepackage{ulem}
\usepackage{multicol}
\usepackage{amsmath}
\usepackage{amssymb}
\usepackage{listings}
\usepackage{tabularx}
\usepackage[super]{nth}
\usepackage{xcolor} 
\usepackage[textsize=footnotesize]{todonotes}
\usepackage{subcaption} 
\usepackage{wrapfig}
\usepackage{float}
\usepackage{multirow} 
\usepackage{tabularx}
\usepackage{booktabs}
\usepackage{rotating}
\usepackage{makecell}
\usepackage{placeins}
\usepackage{etoolbox}
\usepackage{hyperref}
\definecolor{codegreen}{rgb}{0,0.6,0}
\definecolor{backcolour}{rgb}{0.95,0.95,0.92}
\lstdefinestyle{mystyle}{
    backgroundcolor=\color{backcolour},   
    commentstyle=\color{codegreen},
    keywordstyle=\color{blue},
    numberstyle=\tiny\color{gray},
    stringstyle=\color{purple},
    basicstyle=\ttfamily\footnotesize,
    breakatwhitespace=false,         
    breaklines=true,                 
    captionpos=b,                    
    keepspaces=true,                 
    numbers=left,                    
    numbersep=5pt,                  
    showspaces=false,                
    showstringspaces=false,
    showtabs=false,                  
    tabsize=2
}
\newtoggle{NWCHEM}
\settoggle{NWCHEM}{false}

\author{Fusong Ju} 
\affiliation{Microsoft Research AI for Science}
\altaffiliation{Contributed equally to this work}
\author{Xinran Wei} 
\affiliation{Microsoft Research AI for Science}
\altaffiliation{Contributed equally to this work}
\author{Lin Huang}
\affiliation{Microsoft Research AI for Science}
\altaffiliation{Contributed equally to this work}
\author{Andrew J.~Jenkins}
\affiliation{Microsoft Azure Quantum}
\altaffiliation{Contributed equally to this work}
\author{Leo Xia} 
\affiliation{Microsoft Research AI for Science}
\author{Jia Zhang}
\affiliation{Microsoft Research AI for Science}
\author{Jianwei Zhu} 
\affiliation{Microsoft Research AI for Science}
\author{Han Yang} 
\affiliation{Microsoft Research AI for Science}
\author{Bin Shao}
\affiliation{Microsoft Research AI for Science}
\author{Peggy Dai}
\affiliation{Microsoft Research AI for Science}

\author{Ashwin Mayya}
\affiliation{Microsoft Azure Quantum}
\author{Zahra Hooshmand}
\affiliation{Microsoft Azure Quantum}
\author{Alexandra Efimovskaya}
\affiliation{Microsoft Azure Quantum}
\author{Nathan A.~Baker} 
\affiliation{Microsoft Azure Quantum}
\author{Matthias Troyer} 
\affiliation{Microsoft Azure Quantum}
\author{Hongbin Liu}
\email{Hongbin.Liu@microsoft.com}
\affiliation{Microsoft Azure Quantum}

\title[An \textsf{achemso} demo]
  {Acceleration without Disruption: \\DFT Software as a Service}


\begin{document}







\begin{abstract}
Density functional theory (DFT) has been a cornerstone in computational chemistry, physics, and materials science for decades, benefiting from advancements in computational power and theoretical methods. This paper introduces a novel, cloud-native application, Accelerated DFT, which offers an order of magnitude acceleration in DFT simulations. By integrating state-of-the-art cloud infrastructure and redesigning algorithms for graphic processing units (GPUs), Accelerated DFT achieves high-speed calculations without sacrificing accuracy. It provides an accessible and scalable solution for the increasing demands of DFT calculations in scientific communities. The implementation details, examples, and benchmark results illustrate how Accelerated DFT can significantly expedite scientific discovery across various domains.
\end{abstract}


\newpage
\section{Introduction} \label{intro}
Density functional theory (DFT) is a versatile and widely used computational method to study the electronic structure of chemical systems. From OLEDs~\cite{kordt2015modeling,gomez2016design} to drugs~\cite{sulpizi2002applications,tandon2019brief,de2023quantum,cole2016applications,becke2007quantum}, DFT has been crucial for understanding the underlying mechanisms that give rise to specific properties in materials, and for designing new functional materials. Its widespread applicability across various scientific domains has made it one of the most cited theories in the history of physics~\cite{article}. 
Over the years, the accuracy of DFT calculations has been improved by development of various exchange-correlation functionals~\cite{LDA,DFT-Gas,GGA-PW,GGA-Becke,GGA-PBE,metagga1,metagga2,hybrid1,hybrid2,hybrid3,jacob-ladder} and the theory has been extended to study excited states and electronic transitions using time-dependent density functional theory (TDDFT)~\cite{tddft1,tddft2,tddft3,tddft4}. 

The increased utility of DFT is due not only to the development of the theory itself, but also to the dynamic interplay between advancements in programming languages, parallelization techniques, and more significantly, the increasing computational capabilities of hardware over the past five decades. The advancement in computer hardware, especially graphic processing unit (GPUs), has made it possible to reach higher speeds of DFT calculations through hardware-oriented implementation and optimization of DFT codes~\cite{TERACHEM,QUICK,QUICKpaper,BRIANQC,williams2020efficient,williams2023distributed}. 
These developments have dramatically altered the landscape of computational sciences, enabling more complex and large-scale simulations than were previously attainable. 
This increased capability was achieved without implementing approximations such as density fitting~\cite{Dunlap00_37,Ahlrichs04_5119} and linear-scaling techniques~\cite{ONETEP}, which require trade-offs between accuracy and speed. 


Given the increasing complexity of chemistry problems and the demand for rapid digital discovery of new chemicals, there is a critical need for more efficient DFT codes that can perform high-speed calculations with high accuracy. 
This demand is exemplified by the use of machine learning (ML) models for  discovery of new molecules and chemicals, which has shown great potential in accelerating the discovery process by leveraging large-scale DFT datasets~\cite{Butler,m3gnet,smith2020ani,devereux2020extending,schreiner2022transition1x,eastman2023spice}. 
These models can rapidly predict chemical  properties, significantly speeding up high-throughput screening and exploration of chemical space for the discovery of new molecules~\cite{chibattery}. 
The size and accuracy of the datasets used for training these models have a significant impact on performance, further highlighting the need for DFT codes that can provide highly accurate data with high efficiency.  

Here, we present Accelerated DFT, a computational chemistry code for electronic structure calculations of molecules.
It offers a cloud-native, GPU-first approach for the implementation of DFT. 
Due to this approach, Accelerated DFT is able to harness the full computational power offered by contemporary cloud infrastructure and current GPU technologies, achieving an order of magnitude speedup in DFT simulations when compared to other programs using the same GPU or similar CPU cloud resources. 

In this paper, we first provide an overview of the foundational design philosophy behind the key Accelerated DFT algorithms. 
We then detail the redesign and implementation of key components in a conventional DFT workflow. 
Next, we examine the user experience, highlighting the ease of use of the code. We present benchmark results comparing the code's performance against other computational chemistry codes along with its scalability.
Finally, we discuss how this software will accelerate scientific discovery.

\subsection{Cloud-Native Architecture Design} \label{architect}
Our work on hardware-oriented GPU acceleration was inspired by the prior work of others in this space such as TeraChem~\cite{TERACHEM}, QUICK~\cite{QUICK,QUICKpaper}, BrianQC~\cite{BRIANQC}, and GauXC~\cite{williams2020efficient,williams2023distributed}. Over the past decade, rapid advancements in GPU technology have made these processors even more suitable for quantum chemistry calculations due to increased computational power, improved capabilities in double-precision calculations, enhanced parallel-processing capabilities, better memory management, and more mature software frameworks. 
The latest GPUs are often more readily available in the cloud due to their limited availability. Furthermore, cloud computing is becoming increasingly powerful for scientific workloads due to advances in network technology and high-performance computing architectures. For example, some Azure Virtual Machines (VMs) use InfiniBand technology, which provides high-performance, low-latency communication between nodes at scale, a capability that was once exclusive to selected facilities~\cite{azureibrdma}. Thesse advantages have led us to adopt a cloud-native approach in the design of Accelerated DFT. As a result, Accelerated DFT will be offered as a service through an application programming interface (API) using specialized cloud hardware infrastructure, particularly InfiniBand and Remote Direct Memory Access (RDMA), to ensure optimal performance.  

\subsection{GPU-First Algorithm Re-Design} \label{redesign}
To harness the full potential of GPU computing power, we deconstructed  the DFT calculations process, and followed up with a comprehensive redesign and implementation.
Below, we examine the DFT calculations workflow, and then delve into the intricacies of each critical component.

\subsubsection{DFT Calculation Flow} \label{flow}
The DFT computational process is characterized by a sequence of steps designed to ascertain the electronic structure of a given system. Here is a detailed breakdown of these steps:  
\begin{itemize}
    \item \textbf{Initialization of electron density.} The process begins by positing an initial estimate for the electron density, \( \rho(\mathbf{r}) \), which is often derived from superposition of atomic densities or a previous calculation.  
    
    \item \textbf{Formation of the Kohn-Sham Fock matrix.} Based on the initial estimate, the Kohn-Sham Fock matrix can be constructed as: $$\mathbf{F} = \mathbf{H[\rho]} + \mathbf{J[\rho]} + c_{x}\mathbf{K[\rho]} + \mathbf{V}^{xc}[\mathbf{\rho}],$$ where $\mathbf{H}$, $\mathbf{J}$, $\mathbf{K}$ and $\mathbf{V}^{xc}$ are the electron kinetic and external potential, Coulomb, exact exchange, and exchange-correlation (EXC) potential matrices, respectively. 
    
    \item \textbf{Solving the Kohn-Sham equations.} The Kohn-Sham Fock matrix is diagonalized to obtain the eigenvalues and eigenfunctions, known as Kohn-Sham orbitals. This step is performed iteratively, as the orbitals depend on the electron density, which is concurrently being updated.  
    
    \item \textbf{Density update and self-consistency loop.} The electron density $\rho(r)$ is updated based on the occupied Kohn-Sham orbitals. The updated density is then used to recalculate the Fock matrix, and this loop continues until self-consistency is achieved—that is, until the input and output densities are in agreement within a specified tolerance. Direct Inversion in the Iterative Subspace (DIIS) algorithm~\cite{Pulay1980,Pulay1982,li2006energy} is usually used to update the density and accelerate the convergence.
\end{itemize}
Accordingly, it can be deduced that the computational speed of DFT depends on the speed of each iteration and the total number of iterations needed to reach convergence. A better initial guess and more efficient update algorithms, such as DIIS, are helpful in controlling the number of iterations. Many mature methods already exist in this regard and are adopted by most DFT software~\cite{kudin2002black, smidstrup2014improved, garza2012comparison}.

Further acceleration can be realized by completing each iteration faster. In each iteration, the most time-consuming part is the construction of the Fock matrix. For each element in $F_{\mu\nu}$, the computational complexity can be calculated by following formula:
{
\small
\begin{equation}
\label{equ:fock}
F_{\mu\nu} = \underbrace{-\frac{1}{2} \sum_{\mu \nu} P_{\mu \nu} \left( \mu | \nabla^2 | \nu \right)}_{{H}_{\mu\nu}:\ O(1)} + \underbrace{\sum_{\lambda \sigma} P_{\lambda \sigma} \left(\mu \nu | \lambda \sigma \right)}_{{J}_{\mu\nu}:\ O\left(n^2\right)} - \underbrace{c_x\sum_{\lambda \sigma} P_{\lambda \sigma} \left(\mu \lambda | \nu \sigma \right)}_{{K}_{\mu\nu}:\ O\left(n^2\right)} + \underbrace{ \int_{\mathbb{R}^3} \mathrm{d}^3 \mathbf{r}\phi_\mu(\mathbf{r})\frac{\partial E^{xc}[\rho(\mathbf{r})]}{\partial \rho(\mathbf{r})}\phi_\nu(\mathbf{r})}_{{V}_{\mu\nu}^{xc}:\ O(\left|\mathrm{grid size}\right|)},
\end{equation}%
}where $\mathbf{P}$ is the density matrix, $\phi$ is the basis function, and $n$ is the number of basis functions. The most time is consumed in constructing the $\mathbf{J}$ and $\mathbf{K}$ matrices. Since the grid size used for integration of $\mathbf{V}^{xc}$ can be taken as a very large constant ($C$ multiplied by the number of atoms), the overall complexity of this term is $C\cdot O\left(n^3\right)$. For molecular structures with relatively small sizes, the $\mathbf{V}^{xc}$ term dominates the entire Fock matrix. In practice, the $\mathbf{J}$ and $\mathbf{K}$ matrices can be calculated by the Electron Repulsion Integrals (ERIs), which is an analytical integral, and the $\mathbf{V}^{xc}$ matrix is calculated by a numerical integral. We refer to the latter integral as the Exchange-Correlation Energy (EXC) calculation.

In addition to the construction of the Fock matrix, its diagonalization has an $O(n^3)$ complexity. The execution of the DIIS update algorithm also requires a considerable number of matrix operations. 

The primary advantages of Accelerated DFT are its highly efficient algorithms for ERI and EXC calculations, along with efficient implementations of the Fock diagonalization and DIIS on GPUs. In the following section we discuss these algorithms in more detail.

\subsubsection{Key Components Redesign and Implementation} \label{key-components}
\paragraph{\textbf{Electron Repulsion Integrals}}
In Eq.\eqref{equ:fock}, during the construction of $\mathbf{J}$ and $\mathbf{K}$ matrices, the term $\left(\mu\lambda|\nu\sigma\right)$ stands for the following integral:
$$\iint_{\mathbb{R}^3}\mathrm{d}^3\mathbf{r}\mathrm{d}^3\mathbf{r}'\:\frac{\phi_\mu(\mathbf{r})\phi_\lambda(\mathbf{r})\phi_\nu(\mathbf{r}')\phi_\sigma(\mathbf{r}')}{|\mathbf{r}-\mathbf{r}'|}$$ 
For a Gaussian-Type Orbital (GTO) basis set, this integration can be efficiently computed by analytical integral methods, such as HGP~\cite{hgp-head1988method} and Rys~\cite{rys-dupuis1976evaluation}. However, these integration methods pose significant challenges for implementation on GPUs. First, integral screening method that used to work well on CPUs, like Cauchy-Schwarz inequality~\cite{cauchyschwarz1,cauchyschwarz2}, is hard to be implemented on GPUs. This is because inconsistency in the amount of computational data after the screening can greatly reduce parallel efficiency. Second, these methods are based on recursive implementations, which have very complex computational graph dependencies. Moreover, for different angular momentum combinations, there are significant inconsistencies in the computational graph paths. These calculations are vastly different from regular matrix computations, making them highly unfavorable for parallel execution on GPUs.

\begin{figure}[H]
   \centering
    \includegraphics[trim=1cm 2cm 2cm 2cm, width=1\textwidth]{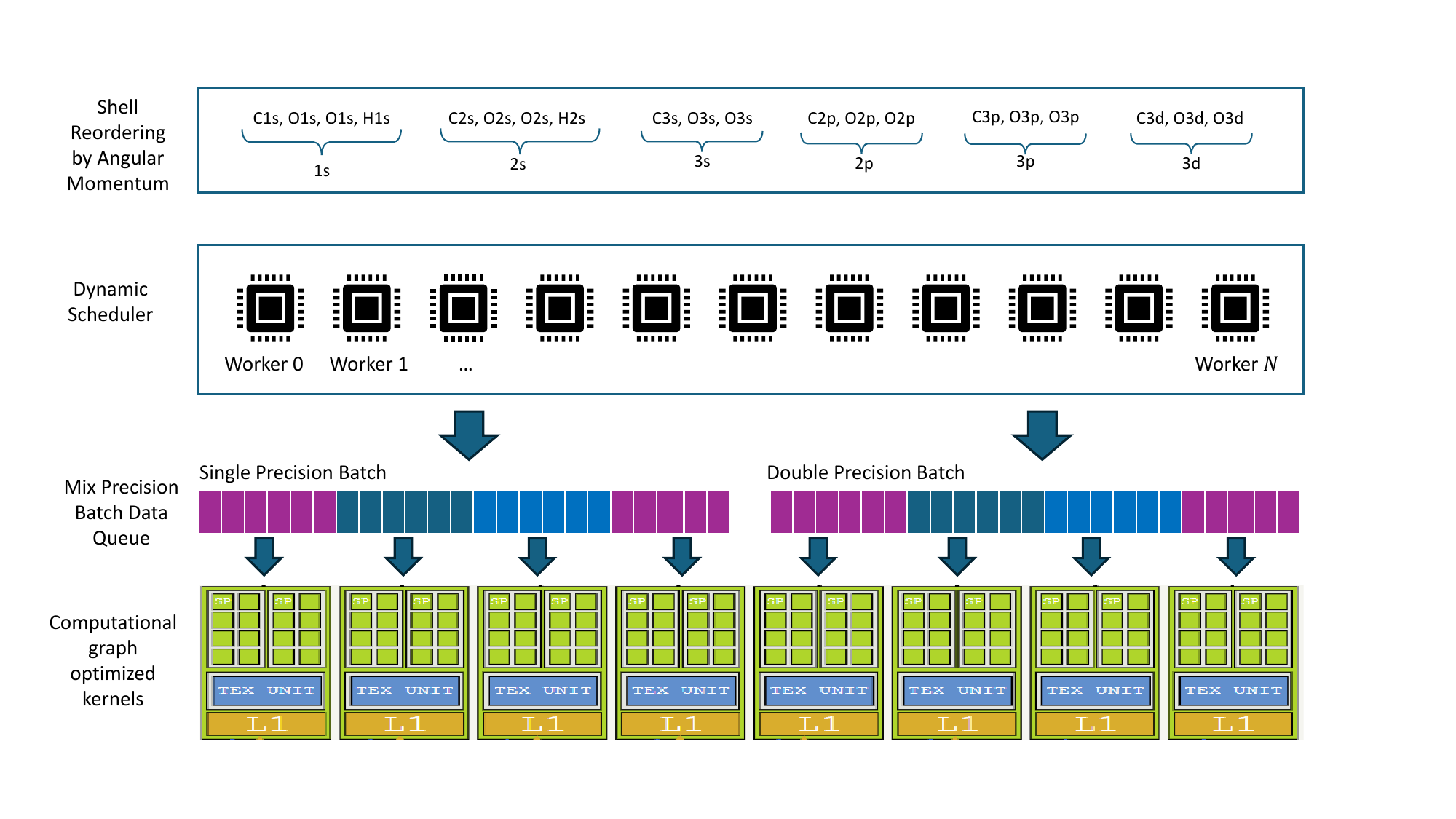}
    \caption{Electronic Repulsive Integral workflow as implemented in the Accelerated DFT code. 
    The workflow starts by reordering the angular momentum and grouping the angular momentum pairings. The pairs are then sent to multiple workers for Cauchy-Schwarz screening. Based on the accuracy, the filtered batches are assigned to mixed precision batch queues. Finally, the batches from each queue are processed by an optimized GPU kernel.}
    \label{fig:eri:flowchart}
\end{figure}

In response to these challenges, we have redesigned the entire integral calculation process in HGP to align with the processing characteristics of GPUs as follows: 
\begin{itemize}
    \item \textbf{Dynamic scheduling of batch data with GPU-friendly angular momentum reordering.} Given that different combinations of angular momentum have completely different recursive computational graphs, we first need to reorder the angular momentum. For this reason, data is prepared in batches based on the combinations of angular momentum. In addition, basis contractions are taken into account in the reordering process, {\textit{i.e.}}, the scheduler treats multiple contractions with the same angular momentum as different shells. For example, all integrals with the angular momentum combination of $(pp|dd)$ are grouped together, while those with $(dd|dd)$ are in another group. In each group, multiple CPU workers perform Cauchy-Schwarz filtering on each integral. The integrals selected for calculations are placed in the batch data to be sent to the GPU. To mitigate the overhead of transferring batch data from CPU memory to GPUs, we utilize multi-queue processing and leverage the asynchronous computing capabilities of the GPU. Specifically, for each GPU, we assign multiple CPU workers with multiple GPU streams. After processing a batch of data, each worker submits it to a stream of GPU for calculation. Then, instead of waiting for the GPU calculation to complete, the worker starts processing the next batch. The GPU automatically processes the computation tasks in a first-in-first-out manner within each stream. By employing multiple workers and streams concurrently on each GPU, we significantly reduce the data processing overhead and achieve GPU utilization rates exceeding 99\%. 
    \item \textbf{GPU kernel code optimized for each type of angular momentum.} Each type of angular momentum combination has a significantly different recursive computational graph, and the GPU is not suitable for handling a large number of branch choices. Therefore, for each type of angular momentum combination, we optimize and determine a unique computational path through code generation, thereby maximizing the operating efficiency of the GPU.
    \item \textbf{A mixed-precision strategy with almost no loss of accuracy.} The efficiency of calculations of different precision on a GPU varies significantly when compared to a CPU. To fully exploit the performance of the GPU, we developed a mixed single and double precision strategy. Specifically, we implemented both double precision and single precision versions for RYS and HGP algorithms, and a job dispatcher to decide how to ``mix'' them. The job dispatcher estimates the upper bound of a shell quartet using the Cauchy-Schwarz inequality, then selects the single precision version if it is enough to obtain the required precision. For example, consider a  shell quartet of $A$, $B$, $C$, $D$ with the estimated upper bound of $S(AB|CD)$. 
    \begin{itemize}
        \item If $S(AB|CD) < \text{TOLERANCE}$ ($10^{-12}$ by default) $\rightarrow$ There will not be dispatch to any algorithm;
        \item If $\text{TOLERANCE} < S(AB|CD) < \text{TOLERANCE}* 10^5$ $\rightarrow$  The single precision version is enough to obtain required precision;
        \item If $S(AB|CD) > \text{TOLERANCE} * 10^5$ $\rightarrow$  The double precision version is necessary.
    \end{itemize}
    After Cauchy-Schwarz filtering, each group of integrals is sent to either a single precision batch or a double precision batch according to the estimated integral precision requirement. When enough data is accumulated in a batch, the corresponding single or double precision GPU kernel is called to perform the integral calculation in parallel. To match the GPU's high computational speed, substantial engineering optimizations have been made to the scheduler to improve the efficiency of CPU worker data preparation. The overall workflow is depicted in Fig.~\ref{fig:eri:flowchart}. \\
\end{itemize}


\begin{figure}[h]
    \centering
    \includegraphics[trim=1.1cm 2.1cm 1.1cm 2.1cm, width=1\textwidth]{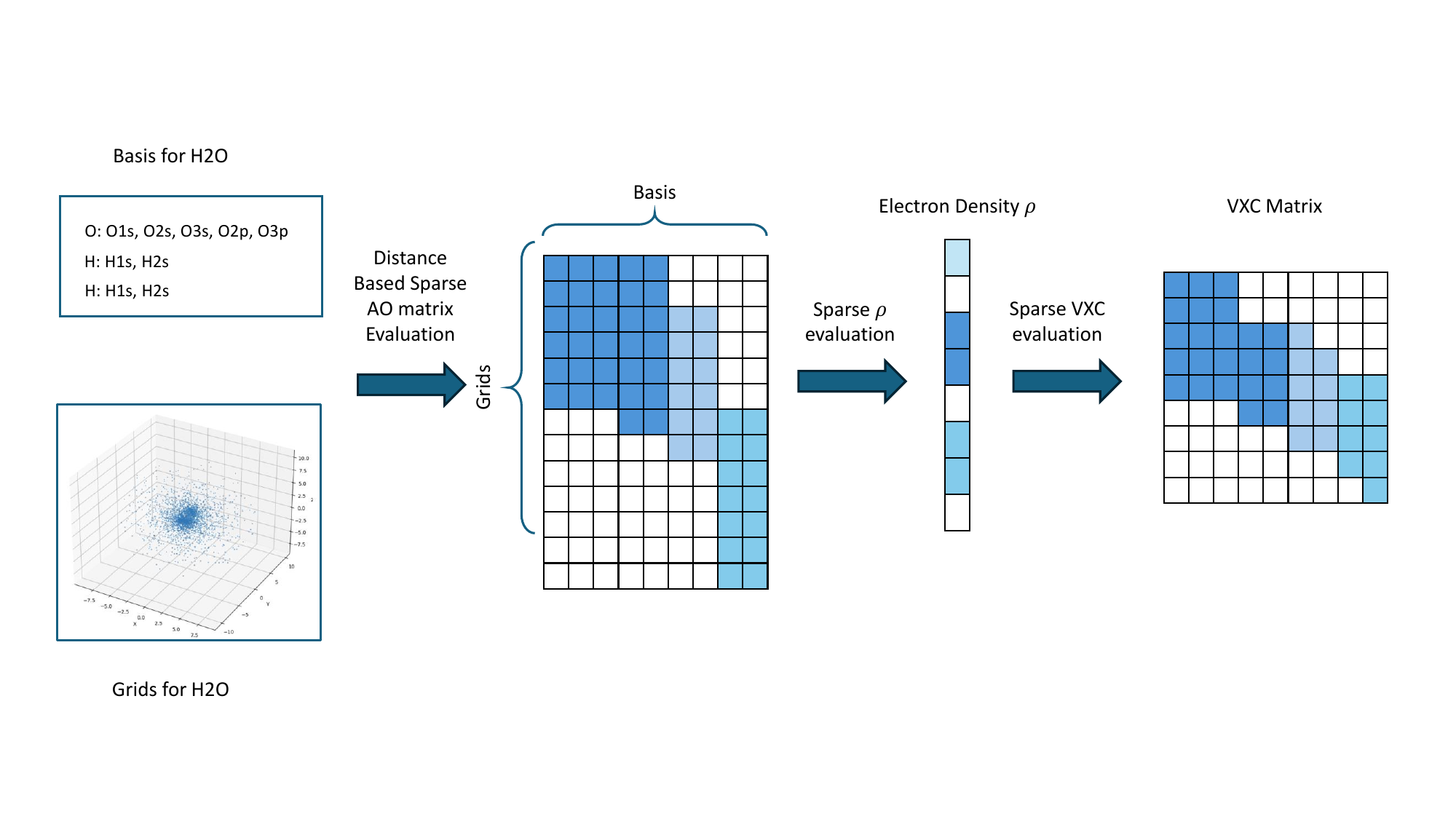}
    \caption{Exchange-correlation energy integral calculation workflow as implemented in Accelerated DFT for $\text{H}_{2}\text{O}$ as an example. For all the basis functions, the sparsity of the AO matrix on the grid points is constructed based on distance from atomic center. Using the obtained sparse AO matrix, the sparse \(\rho\) and \(\mathbf{V}^{xc}\) computation modules are employed to yield the final output.}
    \label{fig:exc:flowchart}
\end{figure}

\paragraph{Exchange-Correlation Energy Integral (EXC)} The calculations of EXC integral are very different from those of ERI. Due to the numerous types and complex forms of exchange-correlation functionals, EXC can usually only be calculated by numerical integration. Compared to ERI, numerical integration is relatively convenient to be implemented on a GPU because the calculations are regular, {\textit{i.e.}}, the same calculation is performed on each grid point, and then the grid points are summed. However, due to the exponential decay characteristics of GTO basis functions, it is possible to ignore the grid points that are far from the atomic center, since the integrated EXC values on those points are small. Taking advantage of this sparsity can dramatically impact the speed of the calculations. However, this sparsity creates challenges for GPU calculations. Although there are some sparse matrix optimization tools, their requirements for sparsity are high, and it is difficult to adapt to the calculation characteristics of EXC. In response to this challenge, we have designed a new sparse optimization algorithm for EXC:
\begin{itemize}
    \item \textbf{Distance-based sparse atomic orbital (AO) matrix evaluation}. Considering the characteristics of the basis function, we set a precision threshold, and for each basis function, we calculate the maximum grid point distance with a value higher than this threshold. For grid points that are farther than this distance, the function is filtered from the data. This saves a considerable amount of computation. To leverage this sparsity, we have optimized the GPU kernels specifically for the computation of the AO matrix.
    \item \textbf{Adaptive sparse block matrix multiplication for $\rho$ and $\mathbf{V}^{xc}$ evaluation}. In order to achieve efficiency gains, it is imperative that we fully exploit the sparsity of the AO matrix during the subsequent computations of the \(\rho\) and $\mathbf{V}^{xc}$ matrices. These computations are heavily reliant on matrix operations. Conventional acceleration techniques for sparse matrices typically require a high degree of sparsity and are not suitable for the weakly sparse nature of the \(\rho\) and \(\mathbf{V}^{xc}\) matrices. Conventional techniques also fail to take advantage of symmetry, thereby impacting efficiency. To address this challenge, we have redesigned a novel matrix computation kernel within the Accelerated DFT framework, to thoroughly capitalize on the specific block-sparsity of the \(\rho\) and \(\mathbf{V}^{xc}\) matrices. We use a fixed block size of 32$\times$32 for our block-sparse strategy in matrix-matrix multiplication during the EXC calculations. The block is considered as zero if all of its 32$\times$32 values are less than $\epsilon$ ($10^{-12}$ by default). Zero blocks are not stored, while all values in non-zero blocks, including zero value, are stored in a dense manner. Conventional kernels, like cuBLAS, cannot be applied on such a data format directly. We implemented a modified version of tiled matrix-matrix multiplication algorithm for this block-sparse format. The overall workflow for EXC calculation is depicted in Fig.~\ref{fig:exc:flowchart}.
\end{itemize}

\paragraph{SCF iteration} As analyzed earlier, the construction of the Fock matrix consumes a lot of time, and once ERI and EXC calculations are moved to the GPU, the diagonalization of the Fock matrix and the DIIS algorithm create new bottlenecks if left in CPU. As a result, the diagonalization and DIIS operations are also implemented in GPU. 
For Fock diagonalization, we use cuSOLVER~\cite{cusolver} for solving, and cuBLAS~\cite{cublas} to complete various matrix operations in DIIS.\\

InfiniBand and RDMA have been used ubiquitously in most of the code implementations. Even though DFT parallelization only involves broadcasting several $N^2$ matrices, and conceptually doesn't rely heavily on a high-speed network. In the case of large-scale DFT calculations, the exchange of matrices takes a non-negligible portion of the total execution time; InfiniBand ensures performance efficiency in those cases. As another example, the dynamic scheduling in ERI and EXC calculations requires frequent updates to a global task counter. We maintain a global shared counter on the rank\#0 process, which needs to be accessed and updated by each worker node. This counter is updated hundreds of thousands of times during a single SCF step in the ERI calculation. The frequent update to the global counter makes the process highly sensitive to data communication latency. RDMA allows us to perform these updates with minimal delay, since it provides the low-latency memory access that is crucial for maintaining the efficiency of our dynamic scheduling algorithm. RDMA uses InfiniBand, which offers both high bandwidth and low-latency communication. This reduces the time spent on data transfer. Additionally, RDMA operations bypass the remote CPU and OS kernel, which leads to lower CPU overhead. This is particularly beneficial in our distributed CPU-GPU heterogeneous computing environment. By leveraging RDMA we can achieve significant performance improvements in our DFT framework. The reduced latency and CPU overhead directly enhance the efficiency of our dynamic scheduling, which leads to faster and more scalable computations. 





\subsection{Minimized Redundancy} \label{redundancy}
Throughout the history of DFT code development, there has been a tendency to create comprehensive packages that aim to meet the diverse needs of researchers using the programs. 
This is because the application of DFT is incredibly varied, with different users focusing on distinct aspects, such as energy, forces, geometry, and other properties. Attempting to encompass all of these facets would result in a significantly larger code base than one designed only for solving the Kohn-Sham equations. This point is illustrated by considering some popular open-source DFT programs. For instance, CP2K v8.2~\cite{CP2K} has more than 900,000 lines of FORTRAN code, and Psi4 v1.3.2~\cite{psi4} has nearly a half million lines of C++ code. The diversity of ways that users consume DFT has resulted in duplicated efforts in utility implementations. For example, almost every DFT code has its own implementation for orbital localization and population analysis, among other tasks. 

To avoid redundancy when developing a new DFT program, it is best practice to evaluate which parts of the design and build require new code, and what features and functions could be built by leveraging existing tools. Our approach in developing a new DFT code is to focus on aspects that significantly enhance performance and provide users with new features and value, rather than attempting to address every possible application or functionality. Thus, Accelerated DFT leverages mature, open-source software codes in the build of non-critical performance parts.

We used open-source software for the following aspects: 
\begin{enumerate}
    \item \textbf{Exchange-correlation functional}. The support of various exchange-correlation functionals is the key feature of any DFT code. 
    LibXC~\cite{libxc} is a mature package that is used in many DFT codes and we employ it to generate the functional form, which is a compute-trivial task. It  also provides agile support for new functionals released yearly. 
    \item \textbf{One-electron integral (with pseudo-potential)}. One-electron integrals contribute to the dominant terms in the Fock matrices, however, their compute complexity is low. We use Libint~\cite{libint} to compute the normal one-electron integral and Libecpint~\cite{libecpint} to compute one-electron pseudo-potential integrals. 
    \item \textbf{Dispersion}. Van der Waals dispersion corrections usually have a very simple analytical form and a very low intensity of compute. These corrections play a critical role in accurately predicting the electronic structure of molecules that include a non-covalent bond. We used simple-dft3 to introduce the D3 correction~\cite{simpledftd3}.  
    \item \textbf{Solvation models}. Solvent can drastically change the electronic structures of molecules. Implicit solvation models like polarizable continuum model (PCM) usually strike a good balance between the cost and accuracy of describing realistic environments. PCM modifies only the one-electron Hamiltonian and thus it does not have a large impact on performance. We use PCMSolver~\cite{pcmsolver} to introduce solvation into the Accelerated DFT code implementation. 
    \item \textbf{Geometry optimizer}. We use geomeTRIC~\cite{geometric} as an optimizer to execute geometry optimization, one of the most commonly performed tasks in any DFT code.  
    \item \textbf{Input/Output schema}. QCElemental~\cite{qcelemental} and QCSchema~\cite{qcschema} were used to standardize and serialize the input and results. 
    \item \textbf{Extensibility}. We use QCEngine's ProgramHarness~\cite{qcengine} to abstract the execution of the DFT code. The program also ensures the future extensibility of the code to seamlessly integrate modified, improved theories for DFT calculations of electronic structure. 
    \item \textbf{Various other utilities}. One of the great challenges in quantum chemistry is the diverging data format of different codes, which makes integration with other tools difficult. To overcome this, our strategy has been to align the data format of molecular orbitals (MOs), density matrices, etc.\ with PySCF~\cite{pyscf}. Therefore, the binary output written by Accelerated DFT can be directly consumed and read in PySCF. Another advantage of this strategy is the generation of additional properties, such as the g-tensor of IR spectra, by calling the corresponding PySCF utilities to re-create the SCF object once the binaries are loaded. 
\end{enumerate}


\section{User Experience} \label{user}
In recent years, cloud computing has shaped a new landscape for HPC. In response, we are introducing a new paradigm in user experience and operational methodology. Below, we discuss key features of the user experience.

From a user perspective, other than the performance of a DFT code, the most important feature is its ease of use. In the contemporary landscape of computational chemistry, running DFT calculations is a process that typically begins with the installation or compilation of specialized software. Due to the computationally intensive nature of DFT, the performance of these programs often hinges on how well they are optimized for the specific hardware on which they run. Scientists typically compile the software directly on their high-performance computing systems to ensure maximum efficiency. This process involves configuring the software to leverage the specific architecture and capabilities of the processors, memory, networks, and sometimes even specialized accelerators such as GPUs.

Once the software is properly installed and optimized, users initiate DFT calculations through command-line interfaces (CLI). The preparation of the input files is a critical step in this process. These files, typically in a text format, must adhere to program-specific syntax and structure. The files contain detailed information about the system of study, including atomic coordinates, basis sets, functionals, and other computational parameters. These files serve as the API of the DFT program. Such design has its origin in historical practices, but it creates a tight restriction on the collocation of user and compute environments. 

\subsubsection{API}
The foundation of a cloud \textcolor {black} {service}, as well as the starting point of the user experience, is the API. The purpose of the API is to distill all operations of the HPC environment to the single element of compute, {\textit{i.e.}}, only the calculation itself is performed on a standard HPC platform and all the extraneous tasks such as preparation of the calculation are executed on any routine computing resource. For example, a common format of a cloud API is the ``Representational State Transfer Application Programming Interface'' (RESTAPI)~\cite{restapi}, which is a set of rules that allows different computer programs to communicate with each other over the internet. This technology enables apps and websites to seamlessly access and share information. 

\subsubsection{Input and Output}
The first step when using the RESTAPI~\cite{restapi} protocol is to decide which data will be exchanged during the communication, {\textit{i.e.}}, the data schema. Here, the data input is defined in a formal but readable syntax so users can focus on scientific content rather than figuring out any program-specific syntax. In modern cloud-app development, JSON is a popular input and output (I/O) data format~\cite{json}. One advantage of the program is the prevalence of the JSON-based data schema,  QCSchema~\cite{qcschema}, advocated by MolSSI to be the I/O standard for quantum chemistry applications. QCSchema has already been reviewed and adapted in the quantum chemistry software community, and is ideal for specifying the I/O parameters for this API application. 

We have made two modifications to the pure QCSchema-based I/O.
First, we acknowledge the likely persistent demand of running DFT calculations on a standard molecular structure format like xyz. Therefore, the Accelerated DFT \textcolor {black} {service} can start the calculations either with a single QCSchema input or with a legacy structure file like xyz, supplemented by a JSON input in the RESTAPI request that specifies the parameters for the DFT calculations. Second, output formats in traditional DFT are not only text, but binary as well ({\textit{e.g.}}, checkpoint files). In the context of a cloud \textcolor {black} {service} and RESTAPI, it is not efficient to always stream binary outputs, which typically range from dozens of megabytes to gigabytes, from the \textcolor {black} {service} to the client. To solve this issue, Accelerated DFT \textcolor {black} {service} integrates with the user's Azure Blob Storage account, which is an object store solution in the cloud. The binary output is stored within a container in the Blob Storage Account~\cite{azureblob}. The user can access the binary files using uniform resource identifiers (URIs) listed in the RESTAPI response. 

\subsubsection{Job Interaction}
Upon defining the API and the I/O for Accelerated DFT \textcolor {black} {service}, a crucial aspect is  determining the methods through which users can interact with the \textcolor {black} {service}. In consideration of traditional practices in DFT usage and the growing demand for automation and scalability, we have implemented a Command Line Interface (CLI) using the Azure CLI. This decision honors the conventional approach familiar to many in the scientific community while also catering to the needs of large-scale computational tasks and automated workflows. 

Furthermore, recognizing the importance of seamless integration with other computational tools and workflows, we have developed a Python Software Development Kit (SDK). This SDK is designed to facilitate easy integration of our DFT \textcolor {black} {service} into a wide range of computational chemistry environments. It allows users to programmatically access DFT calculations, making it simpler to incorporate these tasks into larger, more complex workflows. The Python SDK is particularly advantageous for users who seek to combine DFT calculations with other computational tasks, such as data analysis, machine learning, or other post-processing activities.

The CLI and Python SDK make it convenient for the user to interact with the RESTAPI as well as Azure Blob Storage Account by providing abstractions for uploading input, submitting DFT calculation requests, and downloading output.  For example, the code in line 23 of Listing 1 will interact with the RESTAPI and the Azure Blob Storage account to download the output of the DFT calculation.  

In traditional DFT codes, job management is usually not in the scope of the software. Users rely on external job schedulers or managers like SLURM or PBS to perform tasks such as scheduling and querying jobs for status. However, the DFT cloud \textcolor {black} {service}, presented here, has integrated job management functionality. Jobs can be queried through a web portal, the CLI, or Python SDK. 
List \ref{lst:samplejob} shows an example of how users could submit DFT calculations using the Python SDK and integrate with PySCF for convenient result analysis. \\

\begin{lstlisting}[language=Python, caption={Example of submitting a synchronus DFT calculations using Accelerated DFT service through Python Software Development Kit. As the data format of Accelerated DFT has already been aligned with PySCF, the Accelerated DFT results stored in the QCSchema format can be load directly as a PySCF object. PySCF functions can be utilized directly for result analysis. The Python SDK and CLI abstract the RESTAPI and Azure blob storage interactions
for the user. The ``output = job.get results()'' line in the above sample downloads the data from the Azure blob storage. },label={lst:samplejob}]
# Import the libraries
from azure.quantum import Workspace
from pathlib import Path
# Connect the service with access key
workspace = Workspace.from_connection_string("your_workspace_access_key")
target = workspace.get_targets("microsoft.dft")
# Define parameters
dft_input_params = {
  "tasks": [
    {
      "taskType": "spe", # Single point energy calculations
      "basisSet": { "name": "def2-svp"},
      "xcFunctional": { "name": "m06-2x", "gridLevel": 4 },
      "scf": { "method": "rks", "convergeThreshold": 1e-8 }
    }
  ]
}
# Submit the calculation and get results
job = target.submit(input_data=Path('./ubiquitin.xyz').read_text(),
        input_params=dft_input_params,
        name="ubiquitin-spe")
job.wait_until_completed()  
output = job.get_results() 
# Load result into PySCF DFT object
mol, ks = create_scf_obj(output)
# Use PySCF utilities to compute Milliken charges and dipole moment 
analysis = pyscf.scf.hf.analyze(ks)
\end{lstlisting}

Lines $8-21$ in List \ref{lst:samplejob} could be further compressed into the code snippet in List \ref{lst:qcschemainput} if the input file format is QCSchema.

\begin{lstlisting}[language=Python, caption={Example of submitting a DFT calculation using QCSchema to Accelerated DFT service through Python Software Development Kit.},label={lst:qcschemainput}]
job = target.submit(input_data=Path('./ubiquitin.qcschema').read_text(),
        input_data_format="microsoft.qc-schema.v1"
        name="ubiquitin-spe")
\end{lstlisting}

\section{Tasks} \label{tasks}
A common question for any new DFT code is, ``What can users do with it?'' 
In the design of Accelerated DFT, a key user-experience decision was made to streamline the range of tasks available to users, in contrast to the extensive options provided by traditional DFT packages. Specifically, Accelerated DFT limits its API to accept only five specific tasks: single-point energies (spe), single-point forces (spf), full analytical Hessian calculations (fh), geometry optimizations (go), and Born-Oppenheimer molecular dynamics (bomd)~\cite{helgaker90}. The rationale behind this focused approach is rooted in the fundamental nature of DFT calculations and the evolving needs of the scientific community.

Single-point energy, single-point force, and full Hessians constitute the core and fundamental properties to be evaluated in DFT calculations. These basic calculations lay the groundwork from which a multitude of properties can be derived, such as electron density, population analysis, g-tensors, hyperfine coupling tensors, NMR spectra, IR spectra, and thermochemical properties. Tasks like NMR calculations are identified as separate functionalities in almost all traditional DFT codes; however, in Accelerated DFT, they are viewed as derivative tasks stemming from the three core calculations. This configuration not only simplifies the API but also streamlines the computational process.

Geometry optimization is essentially a serial collection of single-point force calculations. However, it is included as a standalone task due to its widespread use across the entire field of computational chemistry. It remains one of the most commonly executed jobs, necessitating its presence as a primary function in the \textcolor {black} {service}.

The inclusion of BOMD was a forward-looking strategy. Historically, BOMD has seen limited use due to its prohibitively expensive computational cost. However, with advancements in DFT efficiency and the growing interest in using BOMD as a method for sampling potential energy surfaces and chemical spaces, its relevance has increased. This is particularly pertinent in the context of generating data for machine learning potentials and AI applications in chemistry. By supporting BOMD as an individual task, Accelerated DFT aligns itself with these emerging trends and supports the broader scientific community's efforts in these areas. 
Several examples of input files for different tasks are provided in SI.


\section{Benchmarks}
In this section, we present comprehensive benchmarks for the performance of Accelerated DFT in comparison to other well-known codes and in its scalability. For the first case, we compiled a diverse set of molecules, carried out DFT calculations on a single node using Accelerated DFT,
PySCF~\cite{pyscf}, GPU4PySCF~\cite{gpu4pyscf_web}, Psi4~\cite{psi4}, 
\iftoggle{NWCHEM}{TURBOMOLE~\cite{turbomole}, and NWChem~\cite{valiev2010nwchem},}{and TURBOMOLE~\cite{turbomole},}
and provided the accuracy and runtime of Accelerated DFT in comparison to the other codes. 
For the second case, the scaling performance of Accelerated DFT was analyzed with a focus on the parallel efficiency of the code.
This allowed us to identify the optimal balance between time and computational cost, providing a reasonable time-to-solution while analyzing extremely large molecules.

\FloatBarrier
\subsection{Comparative Performance Benchmarking: Accelerated DFT vs. Other Codes} \label{comparative benchmark}
Benchmarking new DFT codes against established ones is crucial for evaluating their performance. In this regard, it is important to assess the performance of the codes on a diverse set of chemical structures, not only to evaluate the robustness of the codes, but also to detect their limitations in handling different chemical systems and to validate their reliability. While many benchmark datasets have been produced and adopted during decades of development of DFT, most of those datasets were created with the purpose of benchmarking the accuracy of new functional development, rather than testing the efficiency of one DFT code versus another~\cite{gmtkn55,mgcdb84,mdb2019}. 

To evaluate the performance of Accelerated DFT in comparison to other available DFT codes, we created a diverse dataset of molecular structures, in terms of both size and chemical composition. 
This dataset comprised 329 molecules from the tmQM~\cite{Balcells2020} and PubChem databases~\cite{Kim2023}. 
The tmQM data allowed access to molecules with catalytically active transition metals, whereas the selected PubChem data contained structures with diverse chemical compositions and large sizes. These molecules contained elements from rows 1--4 of the periodic table and a minimum of 100 atoms, each with at least four unique atom types.
Of the 329 molecules, 110 contained transition metals (28 from the 4th row, 54 from the 5th row, and 28 from the 6th row), and two molecules contained the rare-earth element lanthanum.

For all calculations, the def2-tzvpp basis set\cite{Weigend2005} was used, which contains polarization functions and provides for basis functions with high angular momentum. For example, 110 tests have up to \textit{g} functions, 217 have up to \textit{f} functions, and two have up to \textit{d} functions. For the systems studied here, the choice of def2-tzvpp gives a range of 28--3,913 total basis functions. Moreover, def2-tzvpp implements effective core potential (ECP) in elements heavier than krypton to replace the core electrons. We used ECPs in 85 of the electronic structure calculations in the studied dataset. 
The calculations were performed using both M06-2X~\cite{Zhao2008}, a hybrid meta-GGA exchange-correlation functional, and $\omega$B97x~\cite{chai2008long}, a range-separated hybrid GGA exchange-correlation functional, to evaluate the speedup of Accelerated DFT using different hybrid functionals. The XC numerical integrations were done on an integration grid equivalent or close to the level-4 grid setting in PySCF. A narrow convergence threshold was set to achieve high accuracy by using the ERI screening tolerance of $10^{-12}$ and SCF convergence threshold of $10^{-8}$. Analytical \textbf{J, K} formations were used exclusively throughout all calculations presented in this paper; no density fitting or any other numerical \textbf{J, K} builder was used. 
The design of these benchmark calculations means that existing software to be used for comparison must support the def2-tzvpp basis set and support ECP and high angular momentum basis functions (\textit{f} and \textit{g}) that arise due to this basis set. Additionally, the software must support full analytical evaluation of \textbf{J, K}. Within these limits, we compare Accelerated DFT to a set of codes optimized for CPUs: PySCF (v2.4)~\cite{pyscf}, Psi4 (v1.9)~\cite{psi4}, \iftoggle{NWCHEM}{TURBOMOLE (v7.7)~\cite{turbomole}, and NWChem (v7.2)~\cite{valiev2010nwchem}.}{and TURBOMOLE (v7.7)~\cite{turbomole},} and the GPU-optimized software GPU4PySCF (v0.6.17)~\cite{gpu4pyscf_web}.


For all CPU-based codes, benchmark DFT calculations were carried out on the \href{https://quantum.microsoft.com/en-us/quantum-elements/product-overview}{Microsoft Azure Quantum Elements platform}, using nodes equipped with 120 AMD EPYC 7V12 processor cores with 4 GB of RAM per CPU core. For Accelerated DFT, the calculations were performed on nodes with 4 NVIDIA A100 PCIe GPUs with 80GB memory each, and 96 AMD EPYC 7V13 Milan cores. Since GPU4PySCF does not support multi-GPU computation, GPU4PySCF calculations were performed with 1 NVIDIA A100 PCIe GPUs with 80GB memory and 24 AMD EPYC 7V13 Milan cores.

Within the SCF setting as described above, the convergence of the SCF cycle was obtained for all 329 molecules in the benchmark using Accelerated DFT. For the rest of the codes, the convergence was observed to be different as each of these codes has different default SCF optimization procedures, {\textit{e.g.}}, the number of maximum steps in the SCF cycle, the implementations of DIIS and incremental Fock algorithms, the Fermi level shifting, etc. The purpose of this benchmark is not to compare the convergence efficiency of the SCF cycle, but the efficiency of each step within the SCF cycle. Thus, we limited the comparison between the results of Accelerated DFT and other codes to the converged calculations within the same SCF convergence criteria as described above. 

In comparing the results of Accelerated DFT with other codes used in the performance benchmark analysis, we deemed convergence as our first benchmark test since it demonstrates the robustness and stability of a computational code. For Accelerated DFT, all 329 calculations converged using both M06-2X and $\omega$B97x functionals. However, for the other codes, the convergence was not achieved for all the molecules. For example, convergence was achieved in PySCF for 326 molecules with M06-2X (326 with $\omega$B97x), in Psi4 for 323 (325), in TURBOMOLE for 307 (314), and in GPU4PySCF for 327 (327) with M06-2X ($\omega$B97x) functionals, respectively . For the remainder of the discussion in this section, we compare the accuracy, runtime, and speedup between the converged results only. 

In examining the total energies of individual molecules, the results from any two codes were considered consistent if the errors were below $0.001$ Hartree. Errors above this threshold, indicated a different local minimum of the KS wave function from two codes, resulting in fewer SCF iterations and shorter computation time in one code compared to another, which also made the speedup of one code over another incomparable. Setting this criterion for comparing the performance of Accelerated DFT against other codes led to exclusion of 13 (18) data points from PySCF using M06-2X ($\omega$B97x), 10 (16) from Psi4, 20 (47) from TURBOMOLE, and 14 (19) from GPU4PySCF. The mean absolute error (MAE) was then calculated on the filtered results as presented in Table ~\ref{tab:testset_MAE}.

\begin{table}[H]
\centering
\begin{tabular}{|c|c|c|c|c|c|}
\hline
 & \textbf{Accelerated DFT} & \textbf{PySCF} & \textbf{Psi4} \iftoggle{NWCHEM}{& \textbf{NWChem}}{} & \textbf{TURBOMOLE}\\ \hline
\textbf{PySCF} & 0.028 &  - &  - \iftoggle{NWCHEM}{&  -}{}  & - \\ \hline
\textbf{Psi4} & 0.086 & 0.086 & -  \iftoggle{NWCHEM}{& -}{} & -   \\ \hline
\iftoggle{NWCHEM}{\textbf{NWChem} & 0.130 & 0.130 & 0.170 & - & -   \\ \hline}{}
\textbf{TURBOMOLE} & 0.520 & 0.530 & 0.510 \iftoggle{NWCHEM}{& 0.440}{} & - \\ \hline
\textbf{GPU4PySCF} & 0.028 & 0.000 & 0.086 & 0.530 \\ \hline
\end{tabular}
\caption{Mean absolute error in the total energy of the converged DFT calculations using M06-2X as XC functional.
All numbers are in mHartrees.}
\label{tab:testset_MAE}
\end{table}

When compared to Accelerated DFT, PySCF, GPU4PySCF and Psi4 show an MAE within the same order of magnitude, $10^{-5}$ Hartrees, with the smallest MAE values belonging to PySCF and GPU4PySCF due to their implementation of a grid identical to that of Accelerated DFT.
TURBOMOLE, on the other hand, exhibits an MAE one order of magnitude larger due to differences in its grid implementation compared to Accelerated DFT.

Before we proceed with the analysis of the speedup of Accelerated DFT over other codes, it is essential to underscore two key aspects when evaluating relative speedup between codes. First, the computation time in DFT calculations relies heavily on the setup of the calculations, {\textit{i.e.}}, XC functional, grid level for evaluation of XC, convergence criteria, etc. This directly impacts the speedup of one code over another; the results from one benchmark with a specific calculation setup cannot be generalized to indicate the performance of a code in general. Second,  
any claims about higher efficiency of one code over another should be considered in the scope of performance on a large dataset, {\textit{i.e.}}, in terms of ``mean" speedup, because individual cases may represent outliers and deviate significantly from the mean performance of the code (See Fig. S1).

Using the dataset and the calculation setup as discussed at the beginning of this section, we calculated the mean speedup factor of Accelerated DFT over
\iftoggle{NWCHEM}{PySCF, Psi4, NWChem, and TURBOMOLE,}{PySCF, Psi4, TURBOMOLE,} and GPU4PySCF,
as summarized in Table~\ref{tab:speedups}.
To further highlight the efficiency of Accelerated DFT, pairwise mean speedup factors between the three other quantum chemistry codes in the benchmark are also reported in Table~\ref{tab:speedups}.
Notably, while none of the pairwise comparisons among these established codes yielded a mean speedup factor greater than five, Accelerated DFT significantly outperformed all other codes in the benchmark. 
A schematic of distribution of data for each pairwise comparison is given in the Supporting Information (SI) section, Fig.\ S1-S2.
It is noteworthy that individual cases in Accelerated DFT showed speedup factors several times larger than the mean value, as reported in the speedup range in Table~\ref{tab:speedups}. This reinforces the point made previously on assertions regarding the speedup of one code over another. 

We also compared performance of Accelerated DFT with GPU4PySCF calculations on water clusters of increasing size. Inputs and benchmark timing data for these calculations were taken from the GPU4PySCF website~\cite{gpu4pyscf_web_benchmarks}. 
As above, Accelerated DFT calculations were performed with 4  NVIDIA A100 GPUs and 96 EPYC 7V13 Milan CPU cores, and GPU4PySCF used 1 A100 GPU and 24 EPYC 7V13 Milan CPU cores (a quarter of the node, since GPU4PySCF does not support multi-GPU computing). All timing data is presented in Table \ref{tab_gpu4pyscf_water}. We first note that GPU4PySCF calculations performed on the Azure platform were quicker than those provided by the GPU4PySCF authors – to calculate the speedup factors we use timings from the calculations on the Azure platform in order to provide a fair comparison. 
For small systems,
{\textit{i.e.}} up to 30 atoms, GPU4PySCF and Accelerated DFT show similar performance but, as the system size increases, Accelerated DFT significantly outperforms GPU4PySCF. The largest calculation using 8,201 basis functions showed a speedup factor of 7.83. 
This speedup is less than that seen for the 329 molecule test set because the water clusters contain only light elements, with a single \textit{f} basis function per oxygen, whereas the 329 molecule test set contained 110 molecules with \textit{g} functions and 217 with \textit{f} functions. This demonstrates the enhanced speed of Accelerated DFT when using high angular momentum basis functions, which are needed for accurate calculations. 

In summary, Accelerated DFT clearly exhibits robustness and efficiency. Its substantial speed, when measured against widely used quantum chemistry software programs, highlights its benefits in supporting high-throughput calculations and modeling complex molecular systems.

\begin{table}[H]
\begin{center}
\scriptsize
\iftoggle{NWCHEM}{\begin{tabular}{|cc|ccccc|}}{\begin{tabular}{|cc|cccc|}}
\hline
\multicolumn{6}{|c|}{{\textbf{XC = M06-2X}}}\\ \hline
 & &  \textbf{PySCF} & \textbf{Psi4} & \iftoggle{NWCHEM}{\textbf{NWChem} &}{} \textbf{TURBOMOLE} & \textbf{GPU4PySCF} \\ \hline

\multirow{3}*{\makecell[l]{\textbf{Accelerated} \\ \textbf{DFT}}} & Mean Speedup &  21.82 & 14.86 & \iftoggle{NWCHEM}{23.59 &}{} 17.9 & 12.53 \\  
& Speedup Range & [11.59, 44.13] & [5.50, 37.00] & \iftoggle{NWCHEM}{[2.12, 146.66] &}{} [2.77, 66.11] & [1.59, 60.84] \\ 
& Standard Deviation & 2.69 & 3.59 & \iftoggle{NWCHEM}{14.15 &}{} 8.51 & 6.05 \\ \hline
\multirow{3}*{\textbf{PySCF}} & Mean Speedup  & & 0.68 & \iftoggle{NWCHEM}{1.09 &}{} 0.84 & 0.58 \\ 
 & Speedup Range &  - & [0.35, 1.40] & \iftoggle{NWCHEM}{[0.17, 4.84] &}{} [0.24, 2.48] & [0.14, 3.19] \\ 
 & Standard Deviation &   & 0.14 & \iftoggle{NWCHEM}{0.59 &}{} 0.40 & 0.28 \\ \hline 
\multirow{3}*{\textbf{Psi4}} & Mean Speedup & & & \iftoggle{NWCHEM}{1.60 &}{} 1.37 & 0.95 \\ 
 & Speedup Range & - &- & \iftoggle{NWCHEM}{[0.31, 6.31] &}{} [0.47, 4.52] & [0.27, 7.15] \\ 
 & Standard Deviation &  & & \iftoggle{NWCHEM}{0.76 &}{} 0.89 & 0.68 \\ \hline
 \multirow{3}*{\textbf{TURBOMOLE}} & Mean Speedup & & & & 0.68 \\ 
 & Speedup Range & - &- & - & [0.29, 1.29]\\ 
 & Standard Deviation &  & &  & 0.15 \\ \hline
\iftoggle{NWCHEM}{
\multirow{3}*{\textbf{NWChem}} & Mean Speedup & & & & 1.01 & -\\ 
 & Speedup Range &- &- &-  & [0.16, 4.75] & - \\ 
 & Standard Deviation & &  &  & 0.76 & -\\ \hline
}{}
\multicolumn{6}{|c|}{{\textbf{XC = $\omega$B97x}}}\\ \hline
 & &  \textbf{PySCF} & \textbf{Psi4} & \iftoggle{NWCHEM}{\textbf{NWChem} &}{} \textbf{TURBOMOLE} & \textbf{GPU4PySCF} \\ \hline
\multirow{3}*{\makecell[l]{\textbf{Accelerated} \\ \textbf{DFT}}} & Mean Speedup & 20.15  & 97.14 & 23.27 & 20.39 \\  
& Speedup Range & [7.57, 38.13] & [18.81, 160.76] & [4.14, 93.54] & [2.83, 69.15] \\ 
& Standard Deviation & 3.53  & 32.98 & 12.64 & 8.87 \\ \hline
\multirow{3}*{\textbf{PySCF}} & Mean Speedup  & & 4.70 & 1.20 & 1.07 \\ 
 & Speedup Range &  - & [1.23, 6.65] & [0.55, 3.57] & [0.37, 6.14] \\ 
 & Standard Deviation &   & 1.29 & 0.69 & 0.63 \\ \hline 
\multirow{3}*{\textbf{Psi4}} & Mean Speedup & & & 0.34 & 0.30 \\ 
 & Speedup Range & - &- & [0.12, 1.43] & [0.08, 4.79]\\ 
 & Standard Deviation &  & & 0.30 & 0.37 \\ \hline
 \multirow{3}*{\textbf{TURBOMOLE}} & Mean Speedup & & & & 0.86 \\ 
 & Speedup Range & - &- & - & [0.33, 2.97]\\ 
 & Standard Deviation &  & &  & 0.24 \\ \hline
\end{tabular}
\caption{Comparison of speedup of Accelerated DFT and leading quantum chemistry software for M06-2X and $\omega$B97x XC-functionals on the test set. The mean, range, and standard deviation values are given as the speedup of the code on the left over the code on the right.}
\label{tab:speedups}
\end{center}
\end{table}



\begin{table}[H]
\footnotesize
\centering
\begin{tabular}{cc|cc|cc}
\toprule
 & & \multicolumn{2}{c|}{\textbf{GPU4PySCF}} & \multicolumn{2}{c}{\textbf{Accelerated DFT}} \\ 
 & & \textbf{Official} & \textbf{Azure} \\
 \textbf{NAtoms} & \textbf{NBasis} & \textbf{Time (s)} &  \textbf{Time (s)} & \textbf{Time (s)} & \textbf{Speedup} \\
\midrule
 3 & 59 & 8.02 & 5.67 & 5.51 & 1.03  \\
 15 & 295 & 8.03 & 7.35 & 6.58 & 1.12  \\
 30 & 590  & 17.62 & 15.37 & 8.64 & 1.78  \\
 60 & 1180 & 81.65 & 74.30 & 18.77 & 3.96  \\
 96 & 1888 & 203.23 & 188.19 & 35.35 & 5.32  \\
 141 & 2773 & 454.99 & 421.96 & 71.21 & 5.93  \\
 228 & 4484 & 1482.62 & 1388.79 & 225.40 & 6.16  \\
 300 & 5900 & 2313.47 & 2190.85 & 318.08 & 6.89  \\
 417 & 8201 & 5401.69 & 5229.57 & 668.13 & 7.83  \\
\bottomrule
\end{tabular}
\caption{Comparison of GPU4PySCF and Accelerated DFT performance on a series of water clusters of increasing size (NAtoms= number of atoms, NBasis=number of basis functions). GPU4PySCF `Official times' are taken from the GPU4PySCF website~\cite{gpu4pyscf_web_benchmarks}, and `Azure Time' refers to the same GPU4PySCF calculations run on the Azure platform. Calculation details: M06/def2-tzvpp basis set, 1$e^{-9}$ energy convergence threshold, integration grid =  level 4. Accelerated DFT calculations were performed with 4 A100 GPUs and 96 CPU; GPU4PySCF were performed with 1 A100 GPU and 24 CPU (a quarter of the node) since GPU4PySCF does not support multi-GPU computing.}
\label{tab_gpu4pyscf_water}
\end{table}

\FloatBarrier
\subsection{Scaling Performance Benchmarking: Accelerated DFT on Multiple Nodes} \label{scaling benchmark}

\begin{figure}[h]
    \centering
    \includegraphics[width=\textwidth]{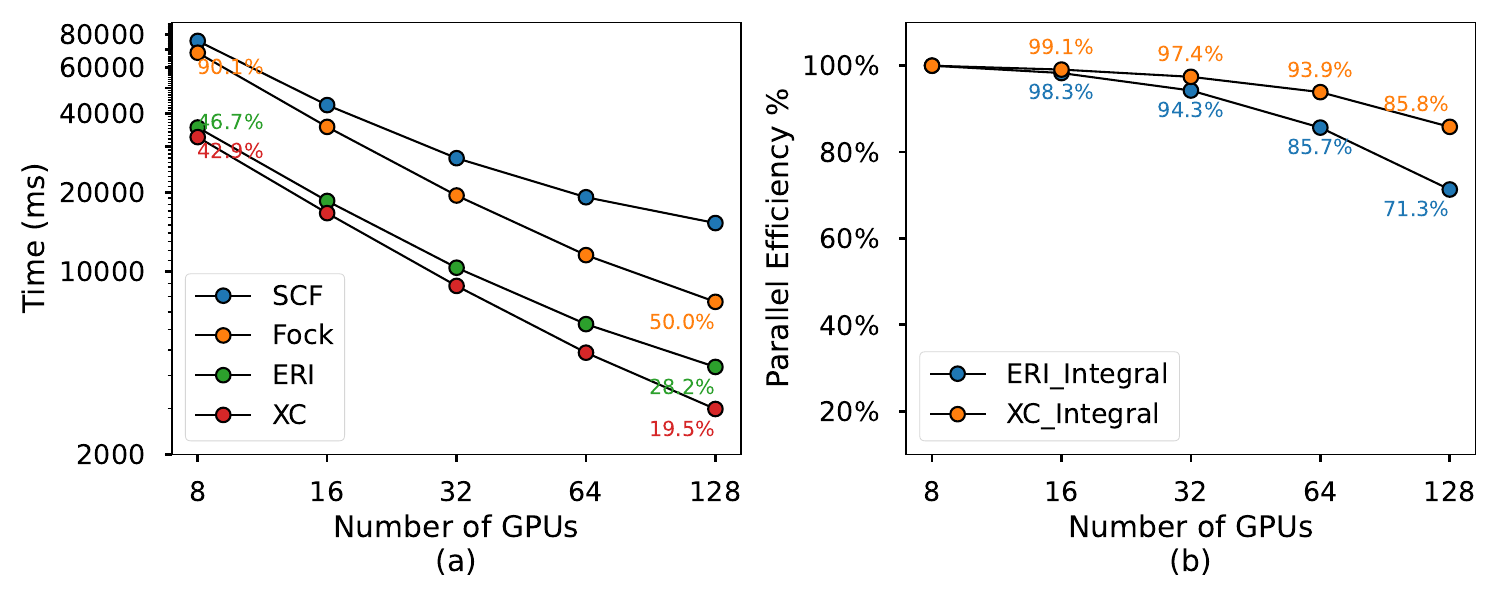}
    \caption{(a) Strong scaling performance of individual components within a single SCF step. Annotations indicate the contribution of each component to the overall SCF step. (b) Parallel efficiency of ERI and XC integral evaluation, where MPI communication time for reducing matrices to rank 0 are excluded.}
    \label{fig:multinode}
\end{figure}

To benchmark the scaling performance of Accelerated DFT, we studied ubiquitin, a large molecule comprising 1,231 atoms~\cite{williams2020efficient}, on 1--16 nodes. 
The calculations were carried out on a cluster of Azure ND A100 v4 series virtual machines. Each node has 8 NVIDIA A100 40G GPUs and 96 AMD EPYC Rome CPU cores. GPU nodes are connected with 200 GB/s NVIDIA Mellanox HDR InfiniBand connections. All GPUs were configured with one MPI process per GPU.

The def2-svp basis set and the M06-2X~\cite{mdb2019} exchange-correlation functional were used for the DFT calculations. All calculations were performed until a convergence of $10^{-7}$ in SCF and $10^{-12}$ in the ERI screening was achieved. For XC integral evaluation, PySCF grid level 4 was implemented, which has 60 radial and 434 angular points for first row elements, 90 radial and 590 angular points for second row elements, and 27,145,282 grid points in total without pruning. All calculations were converged with the exact same energy of $-29,772.5684767$ Hartree after 18 SCF iterations. 

\begin{table}[tbh]
\centering
\begin{tabular}{r|r|rrrr|rrr}
\hline
GPUs & Total & SCF & Fock & ERI & XC & ERI\_MPI (ms) & XC\_MPI (ms) & Bcast\_P (ms) \\
\hline
8   & 1439.4 & 75.6 & 68.2 & 35.3 & 32.5 & 1155.1 & 636.4 & 429.0 \\
16  &  826.4 & 43.0 & 35.5 & 18.5 & 16.7 & 1147.9 & 587.9 & 453.0 \\
32  &  525.2 & 27.0 & 19.5 & 10.3 &  8.8 & 1243.7 & 617.9 & 498.1 \\
64  &  378.2 & 19.2 & 11.5 &  6.3 &  4.9 & 1293.7 & 648.3 & 579.8 \\
128 &  305.5 & 15.3 &  7.6 &  4.3 &  3.0 & 1317.0 & 661.7 & 596.4 \\
\hline
\end{tabular}
\caption{Time of energy calculations (M06-2X/def2-svp) for ubiquitin (1,231 atoms, 11,577 spherical basis functions) on 8--128 A100 GPUs. Total, as shown in second column, is the wall time (in seconds) for completing the DFT job. SCF, Fock, ERI, and XC represent the average time (in seconds) for a single SCF step, Fock matrix construction, ERI, and XC calculations, respectively. ERI\_MPI, XC\_MPI, and Bcast\_P are average MPI communication times (in milliseconds) for reducing $\mathbf{J}$, $\mathbf{K}$ and $\mathbf{V}^{xc}$ matrices and broadcasting density matrix $\mathbf{P}$.}
\label{tab:multinode}
\end{table}

Table~\ref{tab:multinode} presents the time to complete SCF calculations of ubiquitin on 8--128 GPUs. Additionally, it shows the time for completion of individual components within an SCF step to provide a detailed analysis of the scaling performance and parallel efficiency of Accelerated DFT.

For ubiquitin, which comprises 1,231 atoms and 11,577 basis functions in total, the time required for energy calculations decreases consistently as the number of GPUs is increased, indicating strong scaling performance. On 128 GPUs, the SCF calculations can be completed in approximately five minutes. 

The scaling performance and parallel efficiency of individual components within a single SCF step are shown in Fig. \ref{fig:multinode}. 
For distributed ERI and XC integral evaluation, excluding MPI communication time for reducing matrices to rank 0, over $70\%$ PE can be maintained at 128 GPUs. For Fock matrix construction, the parallel efficiency decreases to $56\%$ at 128 GPUs. This leads to a parallel efficiency of about 31\% for SCF cycle at 128 GPUs. This low scaling performance can be mainly attributed to single-GPU calculations on rank 0, including Fock matrix diagonalization ($3.9$ seconds) and DIIS ($2.5$ seconds), which respectively account for $25\%$ and $16\%$ of the time in a single SCF step at 128 GPUs. Nevertheless, the necessity of using InfiniBand can be further echoed as shown in the last row of Table \ref{tab:multinode}; within a single SCF step, the total data exchange can account for approximately $16.8\%$ of the running time. While the ERI and EXC integral computational times can (almost) always be reduced by adding more nodes, the inter-node data exchange starts to represent a significant portion of the total computational time. Keeping the data exchange overhead at a low level is critical to the efficiency of large multi-node calculations.  

These analyses underscore the high efficiency of Accelerated DFT, exemplified by its ability to complete energy calculations for an extremely large system in a very short time. Nevertheless, the observed challenges in optimizing parallel efficiency in certain components of the SCF step highlight the need for further optimization strategies to enhance the overall scalability and efficiency of Accelerated DFT.



\section{Conclusion and Outlook}
We present Accelerated DFT, a GPU-powered cloud-native code with the best performance in the presented benchmarks among selected computational chemistry codes. The high accuracy and efficiency of Accelerated DFT provide an opportunity to substantially expedite research across a wide spectrum of disciplines, from materials science to pharmaceutical development. The acceleration in computational capabilities enables the study of complex molecular systems, which previously required extensive computational resources and time, to be analyzed rapidly and with high precision. 


Beyond traditional quantum chemistry research, Accelerated DFT will expedite AI model developments in the chemistry domain by generating high-quality data more efficiently. In this way, it will catalyze the overall development of new, faster development cycles in chemical research, which will open new frontiers in chemical space and in the discovery of novel molecules such as drugs beyond those available to standard DFT.

\section{Conflict of Interest}
All authors are employees of Microsoft Corporation. Accelerated DFT is a Software as a Service built on Azure, which is Microsoft's cloud platform. Accelerated DFT is a feature offered through Azure Quantum Elements, Microsoft's platform for accelerating scientific discovery.

\begin{acknowledgement}
The Accelerated DFT service has been engineered by a team of developers at Microsoft Azure Quantum. The authors thank Viktor Veis, Xinyi Joffre, Kirill Komissarov, Mahmut Burak Senol, and Peter Shen for their hard work in standing up the service. 
The authors thank Prof. Markus Reiher, Dr. Ajay Panyala, Dr. David Williams-Young, Dr. Jan Unsleber, Dr. Paola Gori Giorgi, Dr. Jan Hermann, Dr. Derk Kooi, Dr. Matthew Horton, and Dr. Rianne van den Berg for fruitful discussions and comments on the manuscript. The authors also thank Deborah Hutchinson  and Michael LeGault for editing and refining the manuscript. 
\end{acknowledgement}

\clearpage
\begin{suppinfo}

Figures \ref{fig:testset_results_speedup_boxplot}-\ref{fig:testset_results_speedup_boxplot_wb97x} shows the pairwise comparison of speedup of Accelerated DFT and leading quantum chemistry software for M06-2X and $\omega$B97x XC-functionals for the 329 molecule test set. 

Figures \ref{fig:testset_results_runtime_grid}-\ref{fig:testset_results_runtime_grid_wb97x} show a pairwise comparison of the compute time of each software for the full test set of molecules. The zero-interception linear regression fit is also plotted; a \textit{y=x} line would indicate equal compute times for the software. Individual points are shaded on a scale from light to dark blue according to the number of basis functions used for that molecule. Accelerated DFT shows a consistently much shorter compute time than all other software.

\renewcommand{\thefigure}{S\arabic{figure}} 
\setcounter{figure}{0}

\FloatBarrier
\begin{figure}[tbhp]
    \centering
    \iftoggle{NWCHEM}{\includegraphics[width=14cm, height=19cm]{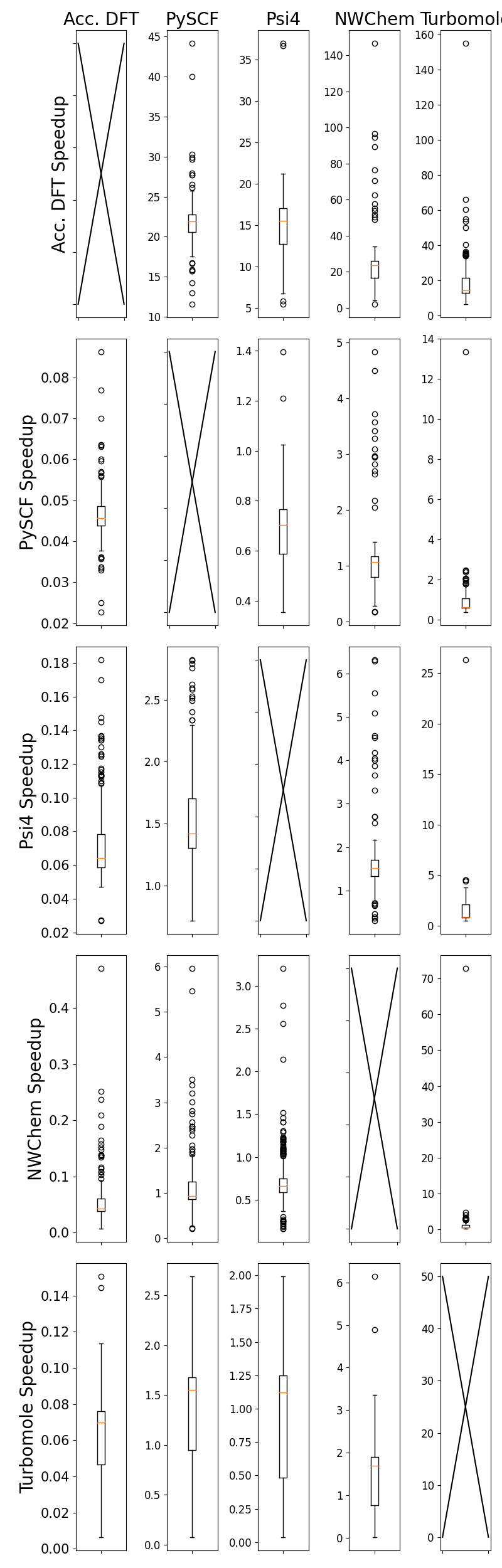}}{\includegraphics[width=14cm, height=20cm]{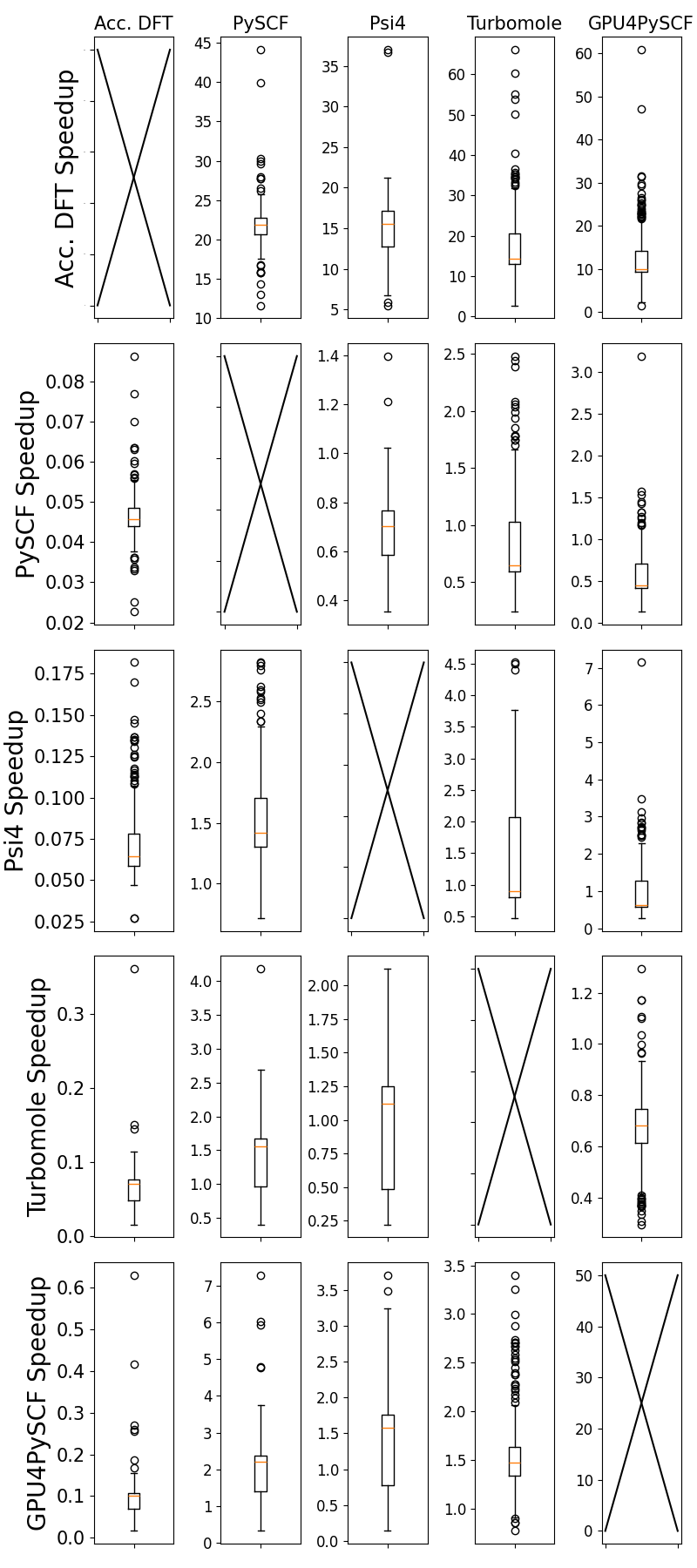}}
    \caption{Boxplot of DFT calculations showing the speedup of Accelerated DFT over leading quantum chemistry software for M06-2X XC functional on the test set. The y-axis is adjusted for each pairwise comparison for visual purposes. The orange line inside each box indicates the median speedup, and the data points outside the whiskers represent the outliers. Accelerated DFT shows a consistent and significantly larger speedup over all other studied software.}
    \label{fig:testset_results_speedup_boxplot}
\end{figure}

\FloatBarrier
\begin{figure}[tbhp]
    \centering
    \includegraphics[width=14cm, height=20cm]{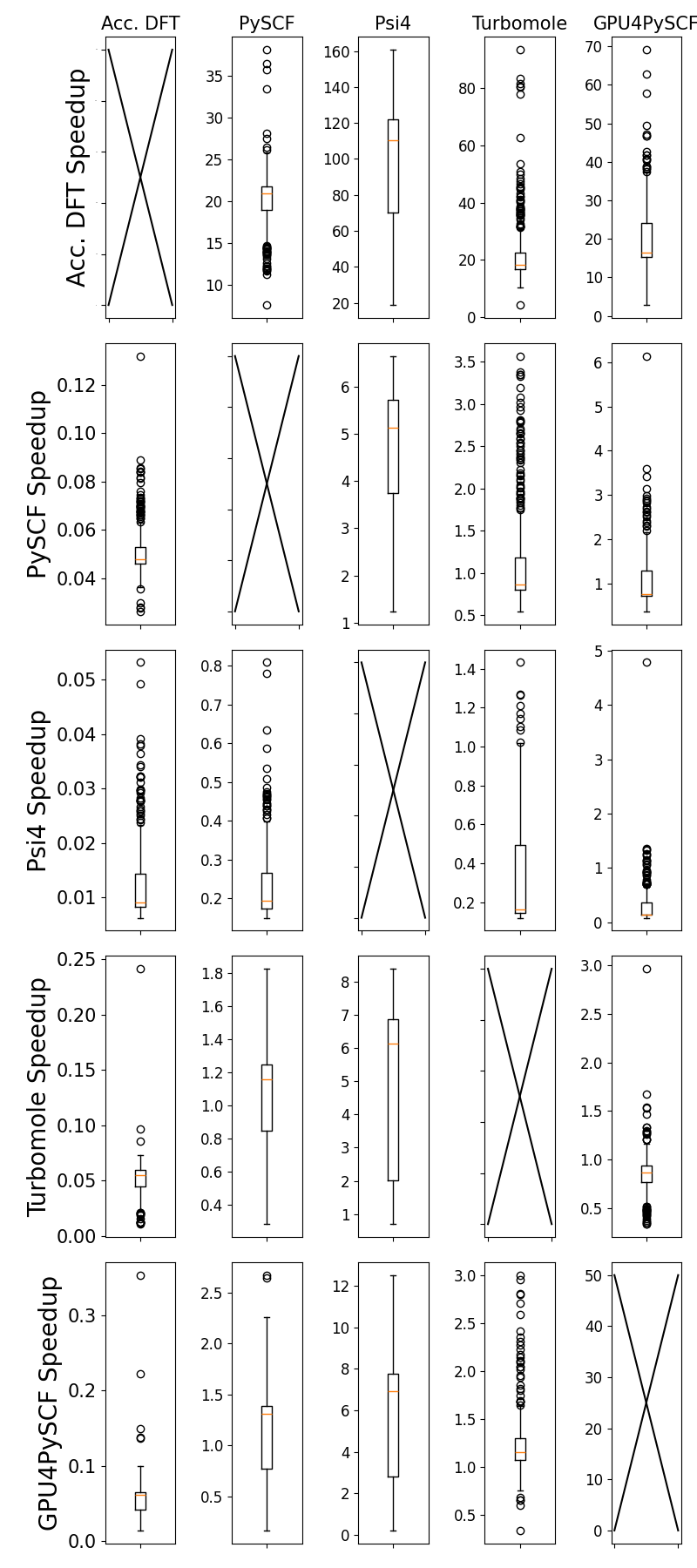}
    \caption{Boxplot of DFT calculations showing the speedup of Accelerated DFT over leading quantum chemistry software for $\omega$B97X XC functional on the test set. The y-axis is adjusted for each pairwise comparison for visual purposes. The orange line inside each box indicates the median speedup, and the data points outside the whiskers represent the outliers. Accelerated DFT shows a consistent and significantly larger speedup over all other studied software.}
    \label{fig:testset_results_speedup_boxplot_wb97x}
\end{figure}

\begin{figure}[tbh]
    \centering
    \iftoggle{NWCHEM}{\includegraphics[width=1.25\linewidth,angle=90]{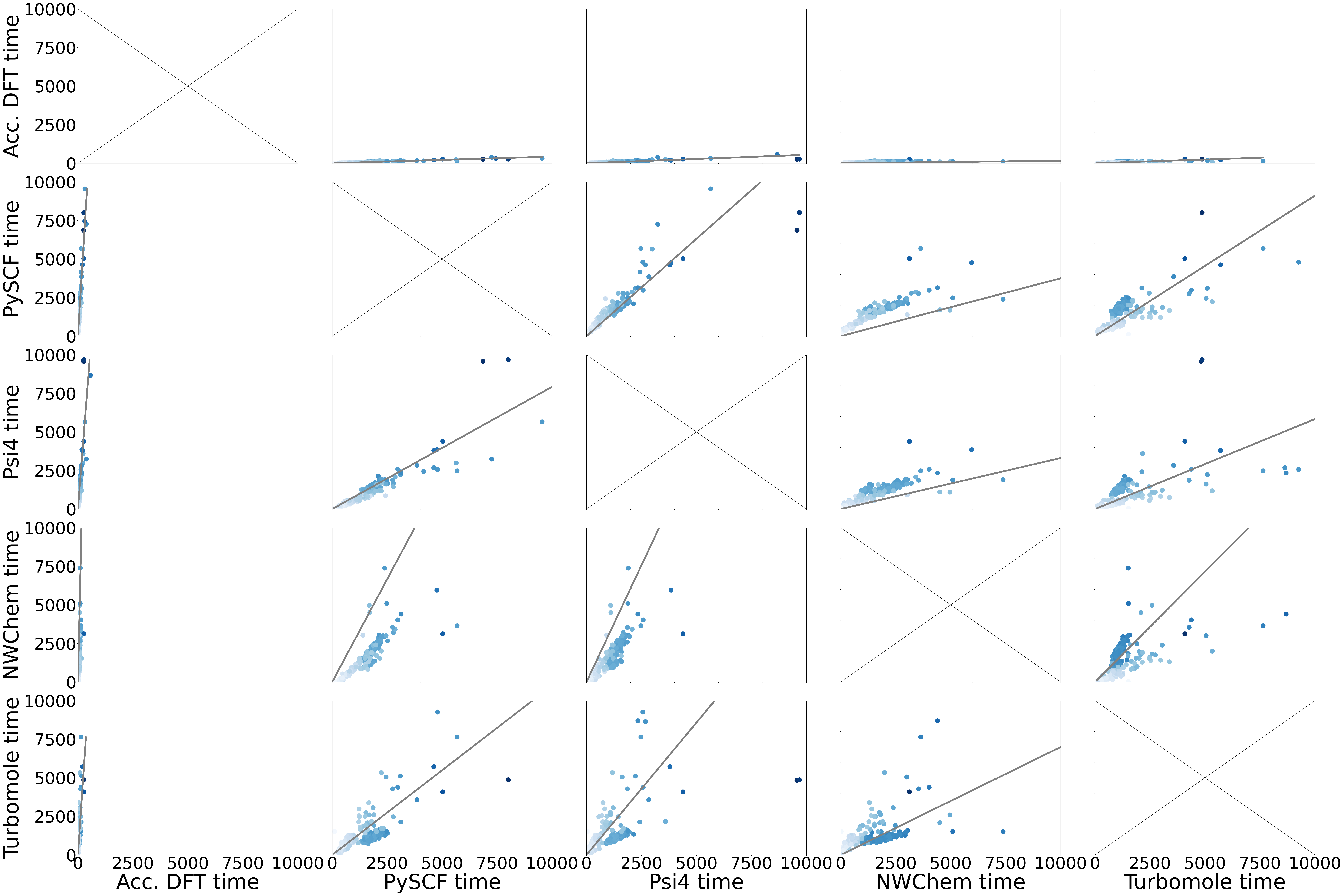}}{\includegraphics[width=1.25\linewidth,angle=90]{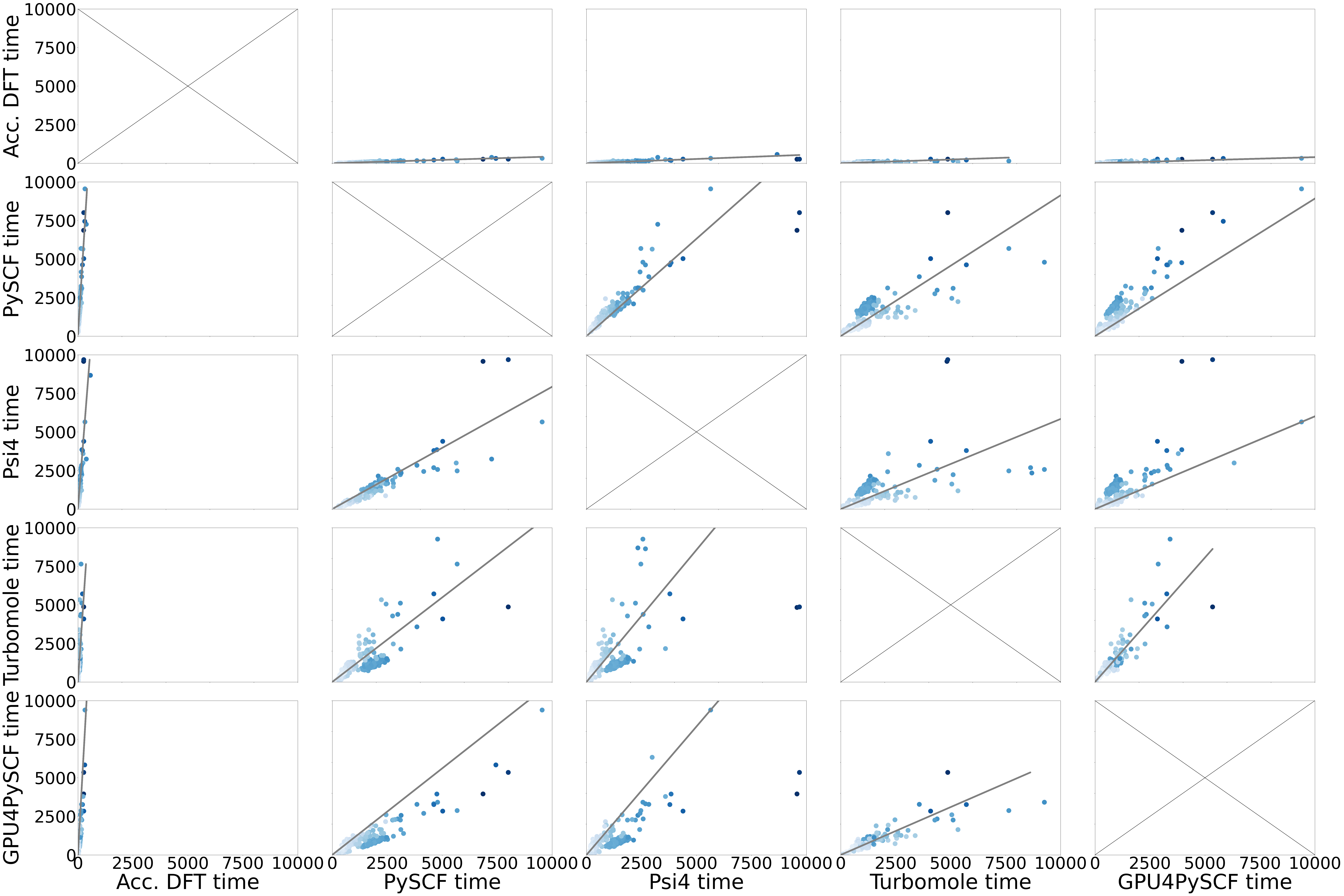}}
    \caption{Parity plots of single point energy runtime between different DFT codes on the 329-molecule dataset using M06-2X. The solid line is the zero-interception linear regression fitting. Individual points are shaded on a scale from light to dark blue according to the size (number of basis functions) of that molecule}
    \label{fig:testset_results_runtime_grid}
\end{figure}

\begin{figure}[tbh]
    \centering
    \includegraphics[width=1.25\linewidth,angle=90]{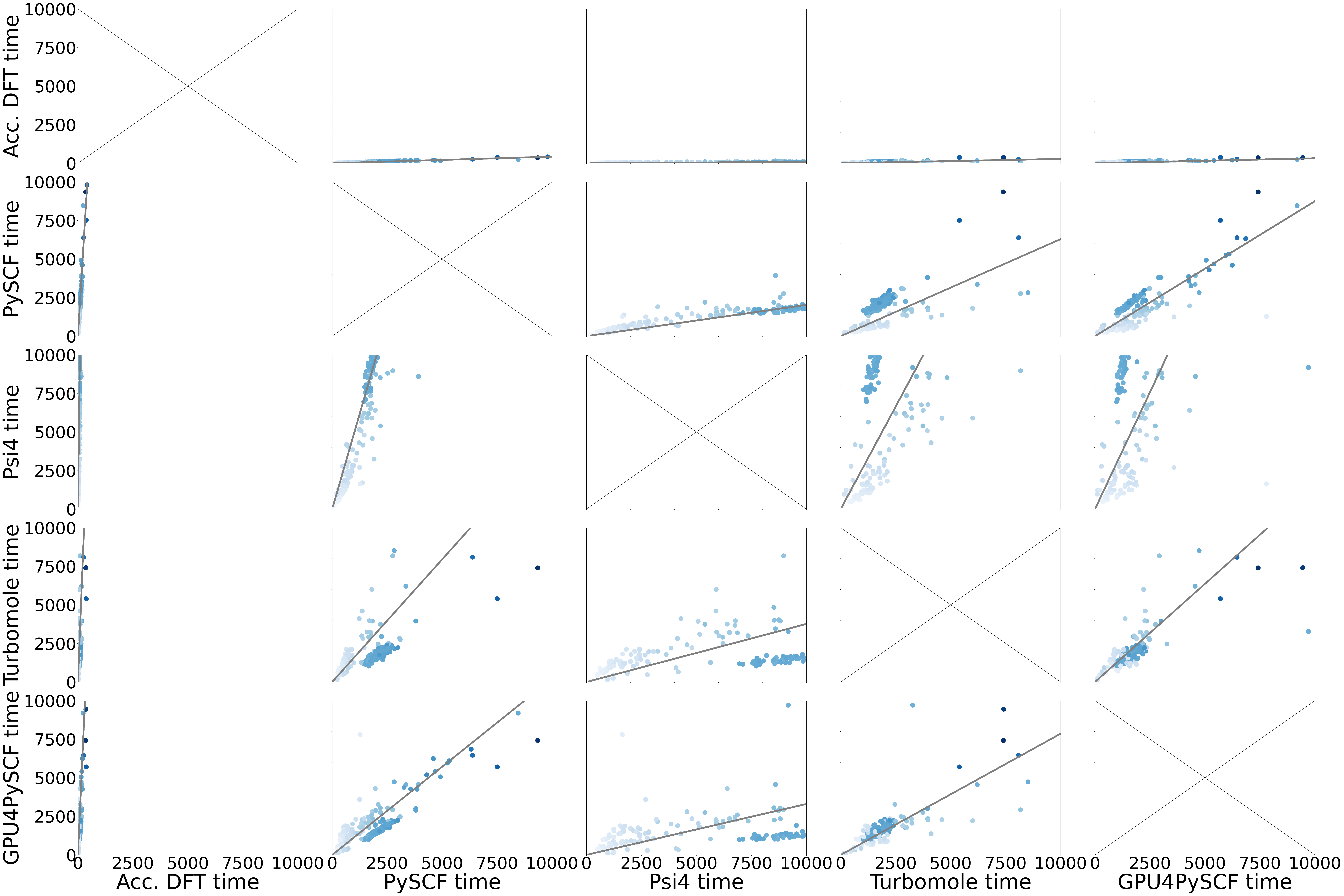}
    \caption{Parity plots of single point energy runtime between different DFT codes on the 329-molecule dataset using $\omega$B97X. The solid line is the zero-interception linear regression fitting. Individual points are shaded on a scale from light to dark blue according to the size (number of basis functions) of that molecule}
    \label{fig:testset_results_runtime_grid_wb97x}
\end{figure}

\clearpage
\renewcommand{\thetable}{S\arabic{table}} 
\setcounter{table}{0}

\newpage

\clearpage
\renewcommand{\thelstlisting}{S\arabic{lstlisting}} 
\setcounter{lstlisting}{0}

List \ref{lst:job1} provides an input example of an Accelerated DFT job of single point energy calculation.
\begin{lstlisting}[language=Python, caption={Example of simple single point energy calculations input.},label={lst:job1}]
dft_input_params = {
     "tasks": [
     {
         "taskType": "spe",
         "molecule": { "charge": 0, "multiplicity": 1 },
         "scf": { "method": "rks", "maxSteps": 100, "convergeThreshold": 1e-8, "requireWaveFunction": True}
         "basisSet": { "name": "def2-svp" },
         "xcFunctional": { "name": "m06-2x", "gridLevel": 4 },
     }
     ]
 }
\end{lstlisting}

Support for PCM solvation models and Grimme-D3 dispersion\cite{Grimme10_154104,Grimme11_1456} are included as additional keywords in the SCF field. Listing \ref{lst:jobgopcmd3} shows an example of defining a geometry optimization in water solvent through IEFPCM and with dispersion correction. 
\begin{lstlisting}[language=Python, caption={Example of adding PCM solvation model and dispersion corrections to DFT calculations.},label={lst:jobgopcmd3}]
 dft_input_params = {
     "tasks": [
     {
         "taskType": "go",
         "molecule": { "charge": 0, "multiplicity": 1 },
         "scf": { "method": "rks", "maxSteps": 100, "convergeThreshold": 1e-8, "dispersion":"d3bj", "pcm": {"solverType": "IEFPCM", "solvent": "water"},"requireWaveFunction": True}
         "basisSet": { "name": "def2-svp" },
         "xcFunctional": { "name": "m06-2x", "gridLevel": 4 },
     }
     ]
 }
\end{lstlisting}

For GO and BOMD tasks, users can supplement with task-specific input parameters. Listing \ref{lst:jobgo} shows how to adjust the convergence criteria for the geometry optimization, changing the energy threshold as well as the maximum and RMS values of the gradient and displacement.
\begin{lstlisting}[language=Python, caption={Example of GO Input.},label={lst:jobgo}]
       "tasks": [
         {
           "taskType": "go",
           "molecule": { "charge": 0, "multiplicity": 1 },
           "basisSet": { "name": "def2-svp"},
           "xcFunctional": { "name": "m06-2x", "gridLevel": 4 },
           "scf": { "method": "rks", "maxSteps": 100, "convergeThreshold": 1e-8 }
           "geometryOptimization": {"convergence_energy": 1e-6, "convergence_grms": 0.0003, "convergence_gmax": 0.00045, "convergence_drms": 0.0012, "convergence_dmax":0.0018 }
         }
       ]
\end{lstlisting}

Listing \ref{lst:jobbomd} shows how to use the ``molecularDynamics" field to set the time propagation, thermostat, timeStep in femtoseconds and temperature in the BOMD calculations. Currently, only the Berendsen thermostat\cite{Berendsen84_3684} is implemented and supported in Accelerated DFT.  The results of the calculations can be directly consumed and rendered in VMD\cite{VMD,VMDweb}. 
\begin{lstlisting}[language=Python, caption={Example BOMD Input.},label={lst:jobbomd}]
       "tasks": [
         {
           "taskType": "bomd",
           "molecule": { "charge": 0, "multiplicity": 1 },
           "basisSet": { "name": "def2-svp"},
           "xcFunctional": { "name": "m06-2x", "gridLevel": 4 },
           "scf": { "method": "rks", "maxSteps": 100, "convergeThreshold": 1e-8 },
           "molecularDynamics":{"steps": 100, "temperature": 300, "timeStep": 1, "thermostat": {"type": "berendsen", "timeSmoothingFactor": 0.05 } }
         }
       ]
\end{lstlisting}

In cases of GO and BOMD tasks, if no task-specific field is given, the default setting (as shown in Listing \ref{lst:jobgo} and Listing \ref{lst:jobbomd}) will be used.

If the requireWavefunction field has been set to True, the MOs will be stored in the Azure storage account and can be retrieved and loaded directly as a PySCF object with the snippet of code in Listing \ref{lst:pyscfobject}. 
\begin{lstlisting}[language=Python, caption={Example creating SCF and mol object in PySCF.},label={lst:pyscfobject}]
     output = job.get_results()
     mol, ks = create_scf_obj(output)
\end{lstlisting}

 At this step, the ``ks" object can be consumed as a normal PySCF KS object for various properties. An example is shown in Listing \ref{lst:dipole}. 
\begin{lstlisting}[language=Python, caption={Example of calculating dipole moment from PySCF DFT object.},label={lst:dipole}]
     # Dipole Moment
     dm = ks.make_rdm1(ks.mo_coeff, ks.mo_occ)
     DipMom = ks.dip_moment(ks.mol, dm, unit='Debye', verbose=3)
     # Molecular Electrostatic Potential
     cubegen.mep(mol, 'file.cube', ks.make_rdm1())
     # CHELPG and RESP atomic charges
     q = chelpg_charges(ks)
     p = resp_charges(ks)
     # NMR chemical shifts
     nmr.RKS(ks).kernel()
     # Polarizability
     polarizability.rks.Polarizability(ks).polarizability()
\end{lstlisting}

If a Hessian `fh' task was carried out in Accelerated DFT, the Hessian can also be loaded and used in the calculation of the vibrational frequencies and thermochemistry information as shown in Listing \ref{lst:freq}
\begin{lstlisting}[language=Python, caption={Example of calculating vibrational frequenices and thermochemistry information in PySCF using the Accelerated DFT Hessian},label={lst:freq}]
     # Load Hessian from QcSchema output
     h = load_qcschema_hessian(output)

     # Compute Vibrational Frequencies
     freq = harmonic_analysis(mol,h)
     dump_normal_mode(mol,freq)

     # Compute Thermochemistry
     thermochem = thermo(ks,freq['freq_au'], 298.15)
\end{lstlisting} 

This information can also be used to compute infrared spectra:
\begin{lstlisting}[language=Python, caption={Example of calculating an infrared spectrum using the Accelerated DFT Hessian},label={lst:IR}]
     # Load Hessian and prepare data
     ks_ir = prepare_ir(ks,output)
     # Compute IR spectrum
     infrared.rhf.kernel_dipderiv(ks_ir)
     ir_intensity = infrared.rhf.kernel_ir(ks_ir)
     # Plot the spectrum
     fig = ks_ir.plot_ir()[0]
     fig.savefig("ir_spectrum.png")
\end{lstlisting}

The following additional information is available from 
\href{https://github.com/microsoft/accelerated-dft}{github.com/microsoft/accelerated-dft}:
\begin{enumerate}
    \item The 329 molecular structures in xyz coordinates reported in the Benchmark section of the paper.
    \item The input settings of running DFT calculations on the 329 molecular structures, including functionals, basis set, grid, and convergence criteria.
    \item The total electronic energies and timings reported from the above mentioned Accelerated DFT calculations.
\end{enumerate}

\end{suppinfo}

\clearpage
\bibliography{achemso-demo}

\end{document}